\documentclass[reprint,amsmath,amssymb,amsfonts,twocolumn,nofootinbib]{revtex4-1}

\usepackage{graphicx}
\usepackage{epsfig}
\usepackage{amsmath,bbm}
\usepackage{amsfonts,amssymb}
\usepackage{times}
\usepackage{dsfont}
\usepackage{enumitem}
\usepackage{array}
\usepackage[dvipsnames]{xcolor}

\newcommand{\ket}[1]{|#1\rangle}
\newcommand{\bra}[1]{\langle#1|}

\newcommand{\expect}[1]{\langle#1\rangle}
\newcommand{\cumulant}[1]{\langle\langle#1\rangle\rangle}

\begin{document}

\title{Cooperative Breakdown of the Oscillator Blockade in the Dicke Model}

\author{Florentin Reiter$^{1,2}$, Thanh Long Nguyen$^{2}$, Jonathan P. Home$^{2}$, and Susanne F. Yelin$^{1,3}$}

\affiliation{$^{1}$Department of Physics, Harvard University, 17 Oxford Street, Cambridge, Massachusetts 02138, USA}
\affiliation{$^{2}$Institute for Quantum Electronics, ETH Z\"{u}rich, Otto-Stern-Weg 1, 8093 Z\"{u}rich, Switzerland}
\affiliation{$^{3}$Department of Physics, University of Connecticut, Storrs, Connecticut 06269, USA}

\date{\today}

\begin{abstract}
The Dicke model, which describes the coupling of an ensemble of spins to a harmonic oscillator, is known for its superradiant phase transition, which can both be observed in the ground state in a purely Hamiltonian setting, as well as in the steady state of an open-system Dicke model with dissipation. We demonstrate that, in addition, the dissipative Dicke model can undergo a second phase transition to a nonstationary phase, characterized by unlimited heating of the harmonic oscillator. Identifying the mechanism of the phase transition and deriving the scaling of the critical coupling with the system size we conclude that the novel phase transition can be understood as a cooperative breakdown of the oscillator blockade which otherwise prevents higher excitation of the system. We discuss an implementation with trapped ions and investigate the role of cooling, by which the breakdown can be suppressed.
\end{abstract}

\maketitle

Spin-boson models, which describe the coherent exchange of excitations between two-level systems and harmonic oscillators, lie at the heart of quantum science.
Most famously, the quantum Dicke model (QDM) describes the interaction of an ensemble of atoms strongly coupled to a single mode of light \cite{Dicke1954, Kirton2019}.
Other than the single-particle quantum Rabi model (QRM), the QDM has not been solved analytically for $N$ particles \cite{Braak2013}. On the contrary, its nonintegrable dynamics renders it an ideal testbed for quantum simulation and many-body quantum physics.
Its most prominent feature is the so-called ``superradiant'' phase transition (SPT): at a critical coupling, the system undergoes a phase transition from a normal to a bright phase, where the spins change their polarization, and the emission into the light mode experiences a collective enhancement, called ``superradiance'' \cite{Hepp1973, Wang1973, Gross1982}.
Over the past decades, a large number of works have studied the QDM, considering entanglement \cite{Schneider2002, Lambert2004, Wolfe2014}, chaos \cite{Emary2003}, implementation in optical lattices \cite{Koch2009}, lasing \cite{Kirton2017}, quantum thermodynamics \cite{Fusco2016}, and even optimization problems \cite{Rotondo2015}.

While the requirement for strong spin-oscillator coupling has made it challenging to implement the QDM directly, emulating it by a cold gas coupled to an optical cavity \cite{Domokos2002, Nagy2010, Baumann2010, Klinder2015} or by Raman transitions \cite{Dimer2007, Zhiqiang2017} have become promising routes.
In addition, trapped ions offer a platform with excellent control for the implementation of spin-boson models \cite{Retzker2007, Pedernales2015, Aedo2018, Lv2018, Lemmer2018} where the QRM and QDM have recently been realized \cite{Lv2018, Safavi2018}.
However, while these models describe a-priori closed systems that are solely governed by a Hamiltonian, real systems exhibit open-system behavior. The resulting nonequilibrium dynamics can enrich the phase diagram and lead to novel phase transitions \cite{Diehl2008, Morrison2008, Verstraete2009, Kessler2012, Lee2014, Genway2014, Bhaseen2014, Zou2014, Carmichael2015, Hannukainen2017, Hwang2018}.
In such dissipative phase transitions, the first of which have recently been demonstrated experimentally \cite{Brennecke2013, Fink2017, Fitzpatrick2017, Rodriguez2017, Fink2017b}, the discontinuity is not in the ground state of the Hamiltonian, but in the steady state of the Liouvillian. It is therefore widely believed that, while the ground state of a closed-system model may be difficult to prepare \cite{Safavi2018}, the driven-dissipative evolution of an open system can freeze out phase transitions such as the SPT in steady state \cite{Buchhold2013, Kirton2016, DallaTorre2016, Larson2017}.
Recently, a new type of dissipative phase transition, the so-called ``breakdown of the photon blockade'' has been studied in the driven Jaynes-Cummings model, giving rise to a highly excited steady-state phase \cite{Carmichael2015, Fink2017}.

\begin{figure}[t]
\centering
\includegraphics[width=\columnwidth]{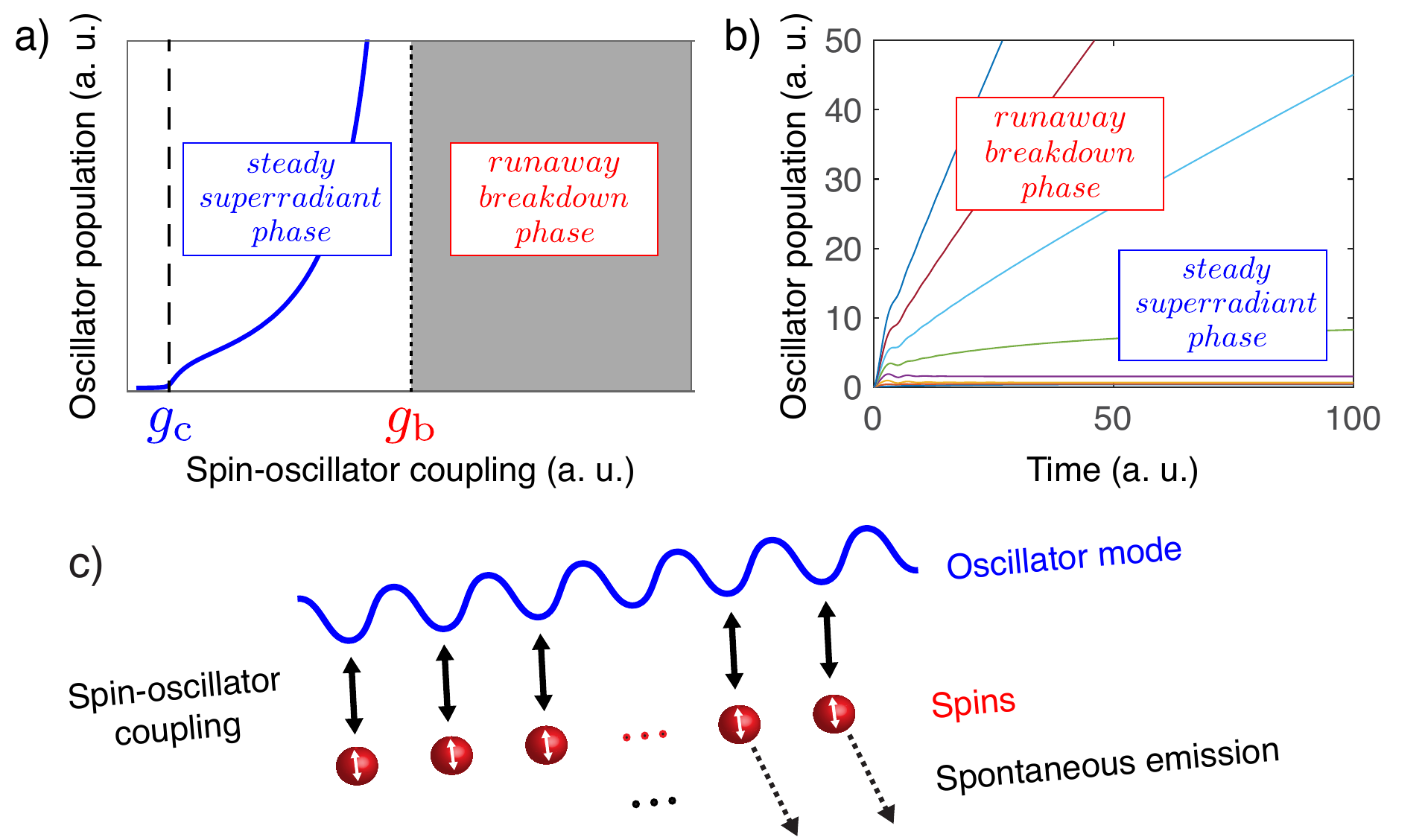}
\caption{
Cooperative breakdown phase transition (BPT) in the Dicke model. (a) Increasing the spin-oscillator coupling causes, beside the well-known superradiant phase transition (dashed boundary at $g_\mathrm{c}$), a BPT in the oscillator (dotted boundary at $g_\mathrm{b}$). While below $g_\mathrm{b}$, the system goes to a steady state, above $g_\mathrm{b}$ it enters a runaway phase characterized by unlimited heating of the harmonic oscillator. Shown are (a) steady-state and (b) dynamical oscillator population obtained from cumulant equations explained in the text.
(c) Schematic of the physical system comprising identical spins coupled to a harmonic oscillator, subjected to local decay by spontaneous emission.
}
\label{fig:breakdown}
\end{figure}

In this Letter, we present a novel dissipative phase transition in the Dicke model which leads to a nonsteady, runaway phase. 
This ``breakdown phase transition'' (BPT) is found in addition to the steady-state superradiant phase transition [see Fig. \ref{fig:breakdown}]. As an underlying mechanism we identify the breakdown of the oscillator blockade by the anti-Jaynes Cummings terms of the QDM.

To investigate the phases of our model beyond mean field we perform a cumulant expansion \cite{Kubo1962} to second order and analytically solve the resulting equations for the steady state. For the SPT, we find a critical coupling that is consistent with previous results. To determine the scaling of the critical coupling for the BPT with the system size we invoke a microscopic model respecting the many-body nature of the effect which cannot be captured by the cumulant expansion. We conclude that the BPT is a cooperative effect, originating from the interplay of critical excitation of strongly dressed many-body resonances and spontaneous emission. We also confirm our theory by numerical simulations.
Addressing the effect of oscillator decay we numerically show that the breakdown can either be maintained or suppressed, depending on the cooling rate.
For the implementation of the model, we consider a system of trapped ions coupled to a motional degree of freedom, which exhibits long coherence times and excellent control allowing for the engineering of the Hamiltonian and the dissipation.
Our work opens up a way to study a new frontier of physics in nonequilibrium phenomena which can be experimentally realized using state-of-the-art platforms.

\textit{Model}. The Hamiltonian of the Dicke model 
is given by
\begin{align}
\hat{H}
&= \frac{\omega_0}{2} \sum_{j=1}^N \hat{\sigma}_z^{(j)} + \omega \hat{a}^\dagger \hat{a} + g \sum_{j=1}^N ( \hat{a}^\dagger + \hat{a} ) \hat{\sigma}_x^{(j)}
%
\label{eq:dicke_hamiltonian}
\end{align}
and describes $N$ identical spins coupled to a harmonic oscillator $\{a^\dagger,a\}$ [see Fig. \ref{fig:breakdown} (c)]. $\omega_0$ is the energy of each spin $j$ (Pauli matrices $\hat{\sigma}_k^{(j)}$, $k \in \{x,y,z\}$, $\hat{\sigma}_\pm = \hat{\sigma}_x^{(j)} \pm i \hat{\sigma}_y^{(j)}$), $\omega$ is the energy of a harmonic oscillator excitation, and $g$ is the coupling constant of the spin-oscillator interaction which is identical for all spins.
The interaction Hamiltonian contains both the energy-conserving, corotating Jaynes-Cummings (JC) terms ($\hat{a}^\dagger \hat{\sigma}_- + \hat{a} \hat{\sigma}_+$), as well as the counterrotating anti-JC terms ($\hat{a}^\dagger \hat{\sigma}_+ + \hat{a} \hat{\sigma}_-$).
As the dominant source of dissipation we consider spontaneous emission described by a jump operator \cite{Agarwal1970,Walls2008}
\begin{align}
\hat{L}_{\gamma,j} = \sqrt{\gamma}  \hat{\sigma}_-^{(j)},
\label{eq:jump}
\end{align}
with a decay rate $\gamma$, acting incoherently on all spins $j$. The discussion of the effect of oscillator decay is deferred to later.

The evolution of operators $\hat{O}$ is governed by the Heisenberg equation of motion
\begin{align}
&\partial_t \expect{\hat{O}} = i \expect{[\hat{H},\hat{O}]} + \expect{\mathcal{D}^\dagger[\hat{L}_{j}](\hat{O})},
\\
&\mathrm{with} ~ \expect{\mathcal{D}^\dagger[\hat{L}_j](\hat{O})}
= -\frac{1}{2} \expect{\hat{L}_{j}^\dagger [\hat{L}_{j}, \hat{O}] + [\hat{O}, \hat{L}_{j}^\dagger] \hat{L}_{j}}
.
\label{eq:dissipator}
\end{align}
To obtain real and compact equations for these observables, we examine the evolution of the Hermitian operators $\hat{\sigma}_x$, $\hat{\sigma}_y$, and $\hat{\sigma}_z$ for the spins, as well as $\hat{q} = \hat{a}^\dagger + \hat{a}$ and $\hat{p} = i (\hat{a}^\dagger - \hat{a})$ for the oscillator. We perform a cumulant expansion beyond mean field \cite{Kubo1962}. This allows us to describe phenomena where correlations are essential. We keep second-order moments (see Ref. \cite{SI} for details), such as the mean oscillator population $\expect{\hat{n}} = \expect{\hat{a}^\dagger \hat{a}} = \expect{\hat{q}^2} + \expect{\hat{p}^2} - \frac{1}{2}$, as well as $\expect{\hat{r}}=\mathrm{Re}\expect{\hat{a}^2}$ and $\expect{\hat{s}}=\mathrm{Im}\expect{\hat{a}^2}$. This results in a closed system of cumulant equations \cite{SI}, describing the dynamics of the expectation values of the spin and the oscillator, $\expect{\hat{\sigma}_z}$ and $\expect{\hat{n}}$. These are nonlinearly coupled to second-order correlations, such as $\expect{\hat{q} \hat{\sigma}_x^{(j)}}$ and $\expect{\hat{\sigma}^{(j)}_x \hat{\sigma}^{(k)}_y}$, for different spins $j \neq k$.
%
We can analytically solve this dynamics for the steady state. The solutions are plotted in Fig. \ref{fig:cumulant}, and further detailed in the Supplementary Material \cite{SI}. Figure \ref{fig:breakdown} also uses data calculated by this method for illustrative purposes.

\begin{figure}[t]
\centering
\includegraphics[width=\columnwidth]{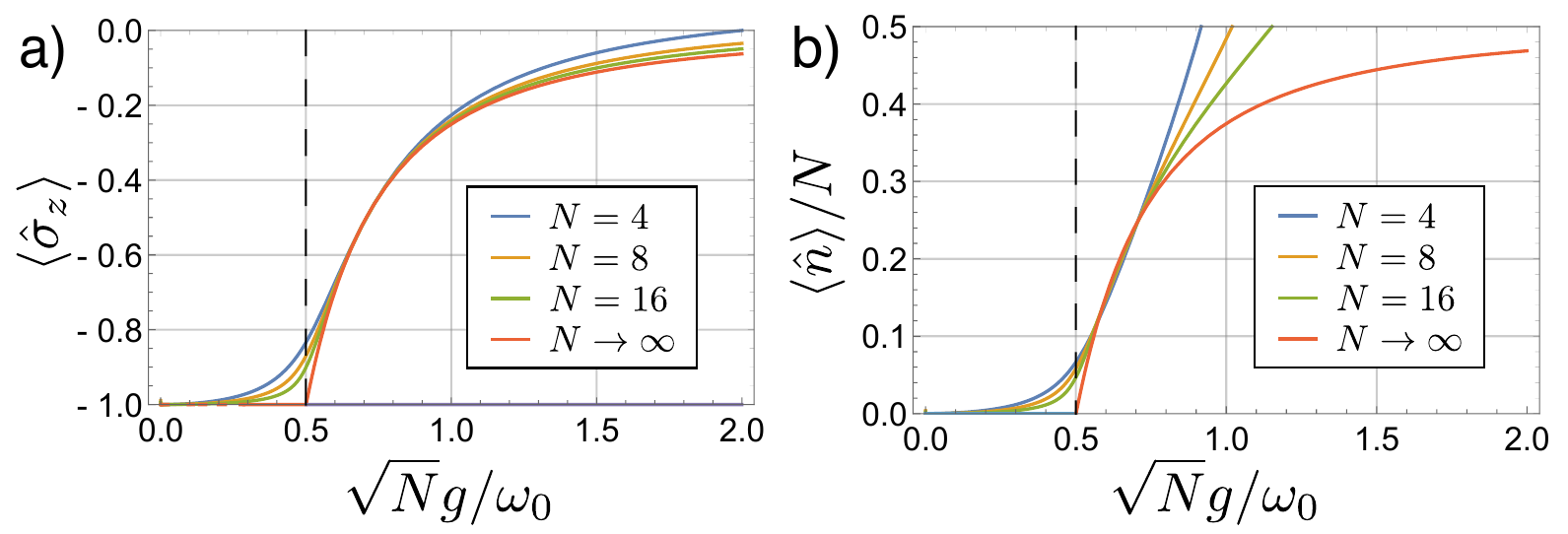}
\caption{
Superradiant phase transition. We plot the analytical steady-state solutions for the polarization of the spins $\expect{\hat{\sigma}_\mathrm{z}}$ in (a), and the renormalized oscillator population $\expect{\hat{n}}/N$ in (b), obtained from the steady-state solution of the cumulant equations, against the renormalized coupling $\sqrt{N}g$. Plots are shown for $\omega = \omega_0$, $\gamma = \omega_0/10$, and particle numbers $N=4,8,16$, as well as $N\rightarrow \infty$. The plots for finite particle number approach the solution for the thermodynamic limit, $N\rightarrow \infty$, where the spins and the harmonic oscillator undergo the superradiant phase transition at a critical coupling $g_\mathrm{c}$ (marked by a dashed line). The cumulant approach used for the data in this figure does not capture the scaling of the breakdown phase transition correctly. An alternative approach will be considered later in the Letter.
}
\label{fig:cumulant}
\end{figure}

\textit{Superradiant phase transition}. For large numbers of particles $N \gg 1$ we find for the steady-state solutions of the polarization and the renormalized oscillator population
\begin{align}
\expect{ \hat{\sigma}_z } &\approx 
\frac{ 1 }{ 8 g^2 } \left( \pm D - 4 N g^2 - \omega \omega_{0, \gamma} \right)
\label{eq:solution:Z}
\\
\expect{ \hat{n} } / N &\approx 
\frac{ \omega_0 }{ 16 N \omega g^2 } \left( \pm D + 4 N g^2 - \omega \omega_{0, \gamma} \right).
\label{eq:solution:n}
\end{align}
Here we have introduced $\omega_{0,\gamma} = ( \omega_0^2 + (\gamma/2)^2 ) / \omega_0$ and a discriminant $D$ which determines the number of solutions,
\begin{align}
D
&\approx \sqrt{ \left( \omega \omega_{0,\gamma} - 4 N g^2 \right)^2 + 16 N g^4 \left( 1 + \frac{\gamma^2}{\omega_0^2} \right) }.
\end{align}
In the thermodynamic limit $N \rightarrow \infty$, we have $D \rightarrow | \omega \omega_{0,\gamma} - 4 N g^2 |$. This gives rise to a nonanalyticity at $D=0$ and discontinuity in $\partial D / \partial g$, respectively,
at a critical coupling \cite{SI},
\begin{align}
g_\mathrm{c} = \sqrt{ \frac{ \omega \omega_{0,\gamma} }{ 4 N } } = \sqrt{ \frac{ \omega \left( \omega_0^2 + (\gamma/2)^2 \right) }{ 4 N \omega_0 } }.
\label{eq:gc}
\end{align}
The nonanalyticity is associated with the second-order superradiant phase transition (SPT) from a normal phase with $\expect{\hat{\sigma}_z}=-1$ and $\expect{\hat{n}}=0$ to a superradiant phase with $\expect{\hat{\sigma}_z}=0$ and $\expect{\hat{n}}/N > 0$ which occurs in both degrees of freedom, spin and oscillator.
These findings are consistent with previous results for open systems \cite{Kirton2016} obtained for $\kappa > 0$ and, for $\gamma=0$, with the closed-system result, $g_\mathrm{c} = \sqrt{ \omega \omega_0 / ( 4 N ) }$ \cite{Hepp1973, Wang1973, Emary2003}.
The criticality of the SPT can be seen from Fig. \ref{fig:cumulant}, where for small $N$ we observe a smooth crossover from the normal to the superradiant phase. For larger $N$, the solutions approach the one in the thermodynamic limit, $N \rightarrow \infty$, which exhibits the nonanalytical SPT at $g_\mathrm{c}$.

\textit{Breakdown phase transition}. In addition to the SPT, in Fig. 1 (a) we observe a second phase transition to a runaway phase at a coupling $g = g_\mathrm{b}$ with $g_\mathrm{b} > g_\mathrm{c}$. In the following, we investigate this phenomenon using the cumulants, as well as perturbative methods, and numerical simulations. For simplicity, we start with $N = 1$, which corresponds to the QRM. Here, there is only a single solution in steady state, 
\begin{align}
\expect{ \hat{\sigma}_z }_\infty &= 
\frac{ g^2 }{ \omega \omega_0 } - 1,
\label{eq:steady:Z}
\\
\expect{ \hat{n} }_\infty &= 
\frac{ \omega^2 + \omega_0 \omega_{0, \gamma} }{ 4 \left( \omega \omega_0 - g^2 \right)} - \frac{ 1 }{ 2 } \left( 1 + \frac{ g^2 }{ \omega^2 } \right).
\label{eq:steady:n}
\end{align}
and no SPT is observed.
However, the mean oscillator population $\expect{\hat{n}}_\infty$ exhibits a pole at a coupling
\begin{align}
g_\mathrm{b,1} = \sqrt{\omega \omega_0}.
\label{eq:gh}
\end{align}
Above $g = g_\mathrm{b,1}$ there is no physical solution ($\expect{\hat{n}}_\infty<0$). This marks a substantial change in the behavior of the harmonic oscillator:
From a dynamical simulation of the cumulant equations in Fig. \ref{fig:breakdown} (b) it can be seen that $\expect{\hat{n}}$ grows to values much higher than in the superradiant phase and keeps increasing over time. At $g_\mathrm{b,1}$ the system thus undergoes a transition to a nonsteady runaway phase characterized by heating of the harmonic oscillator. 
For the BPT, correlations are essential. These are captured by our cumulant approach, while the transition is not obtained from a semiclassical mean-field approach \cite{SI}.

The behavior seen from Fig. \ref{fig:breakdown} partially resembles an effect recently reported as ``breakdown of the photon blockade'', studied in a driven JC model for a single particle \cite{Carmichael2015}, and experimentally observed in a superconducting system \cite{Fink2017}. Here, the anharmonicity of the spin-oscillator interaction in the presence of an excitation prevents higher excitation of the harmonic oscillator by a coherent drive. At the critical coupling $g_\mathrm{b}$, the blockade breaks down, resulting in a highly excited runaway phase.
While in Ref. \cite{Carmichael2015} the breakdown is due to a classical drive and results in a steady state, in the Dicke model at hand the breakdown is caused by the anti-JC part of the spin-oscillator coupling which gives rise to a nonstationary phase.
This can be understood from the fact that the anti-JC couplings $\hat{a}^\dagger \hat{\sigma}_+^{(j)}$ can add two excitations to the system, one to the spins and one to the oscillator. Spontaneous emission then continuously resets the spins, leaving one excitation in the oscillator. This results in a pronounced heating process. Because of the coherent nature of the coupling in the multiparticle Dicke model, the breakdown phenomenon is cooperatively enhanced, as can be understood from the following considerations.

\begin{figure}[t]
\centering
\includegraphics[width=\columnwidth]{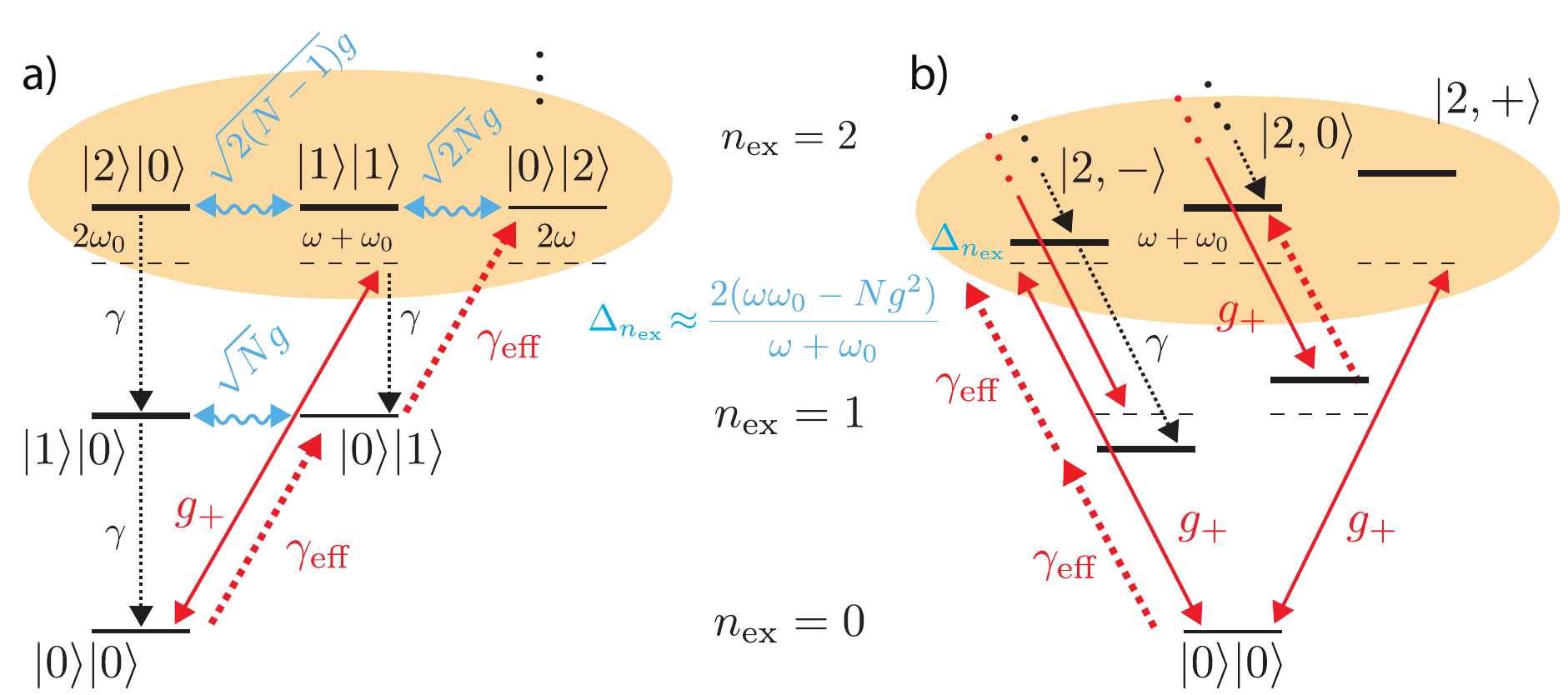}
\caption{
Cooperative breakdown mechanism. In the balanced Dicke model with $\omega_0 \approx\omega$, the JC coupling ($g$) resonantly couples states with the same number of excitations $n_\mathrm{ex}$ (a). As a consequence, these states form dressed states (b).  These are probed by the anti-JC coupling ($g_+$), strongly off resonant and thus blockaded by a detuning $\Delta_{n_\mathrm{ex}} \approx 2 n_\mathrm{ex}(\omega \omega_0 - N g^2)/(\omega + \omega_0)$. For $g_\mathrm{b} \approx \sqrt{\omega \omega_0/N}$ the anti-JC coupling becomes resonant, which results in a breakdown of the oscillator blockade and unlimited heating at a rate $\gamma_\mathrm{eff} \sim \gamma/2$.
}
\label{fig:microscopic}
\end{figure}

\textit{Cooperative breakdown mechanism}. We discuss the breakdown mechanism using the level schemes illustrated in Fig. \ref{fig:microscopic} and the notations $\ket{\psi_\mathrm{spins}} \otimes \ket{\psi_\mathrm{osc}}$ with $\ket{n} = {N \choose n}^{-1/2} (\sum_j \sigma_+^{(j)})^n \ket{00..0}$ for the spins, and $n_\mathrm{ex}$ for the number of excitations.

For a balanced Dicke model with $\omega_0 \approx \omega$, the JC couplings $\hat{a} \hat{\sigma}_+^{(j)}$ and $\hat{a}^\dagger \hat{\sigma}_-^{(j)}$ are on resonance [see Fig. \ref{fig:microscopic} (a)].
On the other hand, the anti-Jaynes-Cummings couplings are detuned by $\omega+\omega_0$.
As a consequence, the states coupled by JC form dressed states, 
which are probed off resonantly by the anti-Jaynes-Cummings couplings. The difference in the detunings of the dressed states in the presence of excitations in the oscillator comprises the blockade mechanism that prevents excitation to higher $n_\mathrm{ex}$.
For example, when starting from the ground state $\ket{0} \ket{0}$, anti-JC excites to $\ket{1} \ket{1}$, which is resonantly coupled by JC to the two other states with $n_\mathrm{ex}=2$, $\ket{0} \ket{2}$ and $\ket{2} \ket{0}$, with coupling constants $\sim \sqrt{2N} g$. The shifts of the dressed states $\ket{2,\pm}$ (see Fig. \ref{fig:microscopic} (b)) are therefore collectively enhanced. For the lower dressed state $\ket{2,-}$ we obtain a detuning $\Delta_{2} \approx 2 ( \omega \omega_0 - N g^2 )/( \omega + \omega_0 )$ \cite{SI}. For suitable coupling $g$, the transition from $\ket{0,0}$ to $\ket{2,-}$  thus becomes resonant with the anti-JC coupling, leading to excitation of $n_\mathrm{ex}=2$ from $n_\mathrm{ex}=0$. The resonance condition $\Delta_{2}=0$ yields the coupling
\begin{align}
g_\mathrm{b} \approx \sqrt{ \frac{\omega \omega_0}{N} } \approx 2 g_\mathrm{c}.
\label{eq:resonance:single}
\end{align}
We obtain the same detunings for higher manifolds and transitions between other dressed states \cite{SI}:
For $N\rightarrow\infty$, the lowest dressed state of each excitation manifold is found to reside at an energy $\lambda_{n_\mathrm{ex},-} \approx n_\mathrm{ex} (\omega \omega_0 - N g^2)/(\omega + \omega_0)$. The resonance condition for excitation with the anti-JC Hamiltonian, $\Delta_{n_\mathrm{ex}} = \lambda_{n_\mathrm{ex},-} - \lambda_{n_\mathrm{ex}-2,-} = 0$, is thus simultaneously fulfilled for all excitation manifolds, resulting in a critical breakdown of the oscillator blockade \cite{SI}.

The excitation by the anti-JC coupling and subsequent reset of the atomic state by spontaneous emission combine to an effective heating process with a rate $\gamma_\mathrm{eff}$.
We estimate $\gamma_\mathrm{eff}$ from $n_\mathrm{ex}=0$ to $n_\mathrm{ex}=1$, as mediated by $n_\mathrm{ex}=2$, using the effective operator formalism \cite{EOF}. For $\omega \approx \omega_0$ we obtain
\begin{align}
\gamma_\mathrm{eff,0 \rightarrow 1} 
= \frac{\gamma N g^2 |\tilde{\omega}_0|^2}{2 |\omega \tilde{\omega}_0 - N g^2|^2},
\label{eq:heating}
\end{align}
with $\tilde{\omega}_0 = \omega_0 - i \gamma/2$.
Excitation of higher manifolds is governed by the respective resonance conditions of their dressed states, $\Delta_{n_\mathrm{ex}}$. The coupling in Eq. \eqref{eq:resonance:single} thus maximizes all effective heating rates $n_\mathrm{ex} \rightarrow n_\mathrm{ex+1}$. Because of saturation, these rates can at most reach $\gamma_\mathrm{eff} \sim \gamma$.
In the breakdown phase, the heating thus becomes comparable to the opposite cooling process: driven by the resonant JC coupling, it also takes place at a rate $\sim \gamma$. The distribution of the oscillator population hence becomes increasingly flat, ultimately approaching an infinite temperature state.

From Eq. \eqref{eq:resonance:single} it can be seen that the breakdown takes place at twice the coupling strength of the SPT. The critical coupling of the BPT is \textit{cooperatively} enhanced by the number of spins, similar to the SPT. This result is, however, different than what would be obtained from the cumulant expansion which yields a single-particle scaling also for $N > 1$ \cite{SI}. We use numerics to resolve the discrepancy and confirm the result in Eq. \eqref{eq:resonance:single}.

\begin{figure}[t]
\centering
\includegraphics[width=\columnwidth]{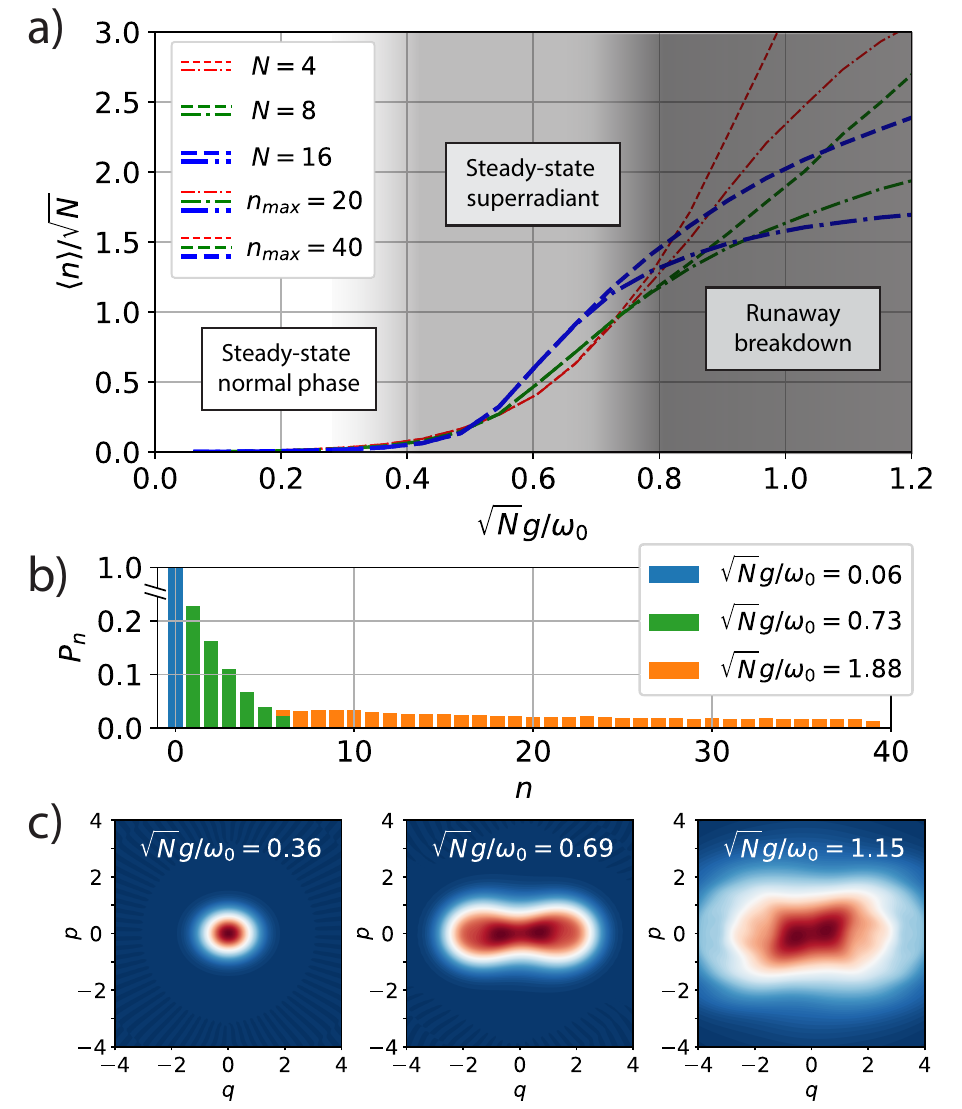}
\caption{
Finite-size numerical simulations.
We plot the mean oscillator population $\langle \hat{n} \rangle$ obtained from integration of the master equation with $N=4,8,16$ spins and a cutoff at $n_\mathrm{max}=20,40$ oscillator quanta. (a) The dependence of $\langle \hat{n} \rangle$ on the cutoff signifies the breakdown of the steady-state superradiant phase.
(b) Population distributions of the harmonic oscillator in the normal phase (blue), superradiant phase (green), and breakdown phase (orange), approaching an infinite temperature state.
Plots are shown for $\omega = \omega_0$, $\gamma = \omega_0/10$, at time $t=1000/\omega_0$. (c) Wigner functions for the harmonic oscillator in the normal phase (left), superradiant phase (center), and breakdown phase (right), simulated for $N=8$ and $n_\mathrm{max}=80$ and shown for $t=1000/\omega_0$.
}
\label{fig:simulation}
\end{figure}

\textit{Numerical simulations}. We simulate the model by numerically solving the dynamics as given by the master equation $\dot{\rho}=-i[H,\rho]+\sum_j\mathcal{D}[L_j](\rho)$. The results are shown in Fig. \ref{fig:simulation} (a) where we plot the oscillator population against the coupling strength. In choosing long enough times we assure that the system gets sufficiently close to the fixed point of the finite-size simulation. We simulate the dynamics for the balanced Dicke model with $\omega = \omega_0$ for $N=4,8,16$ spins and $n_\mathrm{max}=20,40$ oscillator quanta. This allows us to observe the breakdown as a cutoff dependence on the number of simulated oscillator quanta [see Fig \ref{fig:simulation} (b) for the oscillator distribution]. The resulting Hilbert space is a-priori exponentially large in $N$ but the complexity can be reduced to polynomial by exploiting the permutation symmetry of the spins \cite{Kirton2016, Shammah2018}.
The dependence on the cutoff signaling the breakdown of the steady state sets in at around $g \approx (0.8-1)\omega_0/\sqrt{N}$, for $\omega = \omega_0$.
This scaling with the number of the particles confirms the cooperative nature of the BPT.

In Fig. \ref{fig:simulation} (c) we plot the Wigner function for the harmonic oscillator in the normal, superradiant, and breakdown phase. The superradiant phase transition is observable as a separation of the single-peak distribution in the normal phase into two peaks in the superradiant phase, which is a consequence of spontaneous symmetry breaking \cite{Kirton2016}. The breakdown phase transition is accompanied by heating of the oscillator and thus results in an inflation of the distribution.

\textit{Implementation}. 
The direct realization of the Dicke Hamiltonian is challenging as it requires a very strong spin-oscillator coupling $g$, comparable to the energies $\omega, \omega_0$. Otherwise, for $\omega, \omega_0 \gg g$, the anti-Jaynes-Cumming terms can be neglected by the rotating wave approximation. Both JC and anti-JC terms can, however, be implemented as \textit{effective} couplings \cite{Nagy2010, Baumann2010, Dimer2007, Zhiqiang2017}. This has been realized by stimulated Raman couplings between two ground levels of the atoms \cite{Dimer2007, Zhiqiang2017} and by coupling the atomic transition to motional degrees of freedom of a BEC \cite{Nagy2010, Baumann2010}. We consider a realization based on trapped ions, where the (anti-) JC terms are implemented using Raman sideband drives \cite{Meekhof1996, Kienzler2015, Pedernales2015, Lv2018, Aedo2018}.

Our recipe for the simulation of the driven-disspative QDM extends those for the closed-system QRM and QDM in Refs. \cite{Pedernales2015, Lv2018, Aedo2018}. The spins are mapped to internal transitions and the harmonic oscillator is comprised by a motional mode. Two sideband couplings are combined to the spin-oscillator interaction and their detunings yield the energy terms \cite{SI}. The dissipation is realized by an optical pumping laser from $\ket{1}$ to a short-lived level $\ket{e}$ which rapidly decays to $\ket{0}$, leading to the desired jump operator in Eq. \eqref{eq:jump} with the possibility to tune the decay rate $\gamma$.

\textit{Cooling}. Cooling is required to get the motion of the ion crystal close to the ground state \cite{Ballance2015, Negnevitsky2018}. It is formally equivalent to oscillator decay described by a jump operator $\hat{L}_\kappa = \sqrt{\kappa} \hat{a}$ with a decay rate $\kappa$. The presence of a significant cooling process suppresses the breakdown \cite{SI}:
For small cooling rates $\kappa$, the critical coupling for the BPT remains at values $g_\mathrm{b} \approx 2 g_\mathrm{c}$. As the cooling rate is increased, the breakdown is shifted to higher couplings $g_\mathrm{b} \gg g_\mathrm{c}$, way above the SPT. Hence, for sufficiently large cooling rates $\kappa$, the BPT does not occur. This explains why the breakdown has not been observed in cavity systems \cite{Baumann2010, Brennecke2013, Klinder2015}. It should, however, be observable in systems of trapped ions, where the lifetime of the harmonic oscillator comprised by the mechanical motion can be extremely long.

\textit{Conclusions and outlook}. We have investigated the phases of a driven-dissipative Dicke model. Here we find, in addition to the well-known superradiant phase transition (SPT), a novel breakdown phase transition (BPT) to a nonstationary phase of the harmonic oscillator. Analyzing the underlying microscopic processes, we conclude that this phase transition is driven by a cooperative breakdown of the oscillator blockade.
As opposed to the behavior of the SPT, the BPT is not of the mean-field type, but driven by correlations, as is captured by our cumulant description. Our physical explanation of the breakdown relies on the quantized nature of the spins and the oscillator. In addition, dissipation is essential to observe the BPT, which is different from the SPT that also occurs in a closed system. We conclude that the BPT is a nonequilibrium quantum phase transition.
For the implementation, we consider a system of trapped ions and discuss the realization of the couplings. The possibility to add cooling of the harmonic oscillator allows us to observe the Dicke phase transition in steady state while suppressing the breakdown.
Our work enriches the well-established Dicke model by an additional, fundamentally different phase. It enables demonstrations that go beyond steady-state phase transitions, opening up new prospects in nonequilibrium many-body quantum physics.
A particularly interesting direction may be given by the possibility to observe the breakdown phase transition in systems of higher spin, such as multi-level systems of trapped ions \cite{Senko2015}.
Beyond fundamental interest, the cooperative breakdown may be useful as a resource for phase-transition-enhanced sensing schemes \cite{Raghunandan2017, FernandezLorenzo2017, Macieszczak2016}.

\textit{Acknowledgments}. We acknowledge discussions with Ephraim Shahmoon, Yulia Shchadilova, Eugene Demler, Jamir Marino, Shengtao Wang, Julian Leonard, Johannes Fink, and Tony Lee. This work was supported by the Swiss National Science Foundation (SNSF) through the National Centre of Competence in Research - Quantum Science and Technology (NCCR QSIT) grant 51NF40–160591. F.R. acknowledges financial support by a Feodor-Lynen fellowship from the Alexander von Humboldt-Foundation and from the Swiss National Science Foundation (Ambizione grant no. PZ00P2$\_$186040). S.F.Y. thanks the NSF for support via the CUA PFC.

\bibliography{references} 

\onecolumngrid

\clearpage

\section*{Supplementary Information}

In this Supplementary Information to the Letter ``Cooperative Breakdown of the Oscillator Blockade in the Dicke Model'' we present the cumulant expansion leading to the equations and their solutions in Sec. \ref{app:cumulants}, a theory based on dressed states demonstrating criticality in Sec. \ref{app:dressed}, and the implementation of the driven-dissipative Dicke model with trapped ions in Sec. \ref{app:implementation}.

\section{Cumulant expansion} \label{app:cumulants}

The time evolution of the driven-dissipative system represented by the density operator $\rho$ is governed by a master equation of Lindblad form
\begin{align}
\frac{d\rho}{dt} = \mathcal{L}(\rho) = -i [\hat{H}, \rho] + \sum_k \mathcal{D}[\hat{L}_k](\rho).
\label{eq:master1}
\end{align}
The Liouvillian $\mathcal{L}(\rho)$ contains the Hamiltonian $\hat{H}$ and dissipators
\begin{align}
\mathcal{D}[\hat{L}_k](\rho) = \hat{L}_k \rho \hat{L}_k^\dagger - \frac{1}{2}(\hat{L}_k^\dagger \hat{L}_k \rho + \rho \hat{L}_k^\dagger \hat{L}_k).
\label{eq:master}
\end{align}
for each jump operator $\hat{L}_k$.
\\

In the Heisenberg picture, the time evolution of the expectation value of a time-independent operator $\hat{O}$ is described by
\begin{align}
\frac{d}{dt} \expect{\hat{O}} = \frac{d}{dt} \mathrm{Tr}(\rho \hat{O}) = \mathrm{Tr}\left(\frac{d \rho}{dt} \hat{O}\right).
\end{align}
With Eq. \eqref{eq:master} this can be written as
\begin{align}
\frac{d}{dt} \expect{\hat{O}} = i \expect{[\hat{H},\hat{O}]} + \sum_k \expect{\mathcal{D^\dagger}[\hat{L}_k](\hat{O})},
\end{align}
where the dissipative part is given by
\begin{align}
\expect{\mathcal{D}^\dagger[L_k](\hat{O})} = \expect{\hat{L}_k^\dagger \hat{O} \hat{L}_k - \frac{1}{2}(\hat{L}_k^\dagger \hat{L}_k \hat{O} + \hat{O} \hat{L}_k^\dagger \hat{L}_k)}
= -\frac{1}{2} \expect{\hat{L}_k^\dagger [\hat{L}_k, \hat{O}] + [\hat{O}, \hat{L}_k^\dagger] \hat{L}_k}.
\end{align}

The time evolution of an expectation value, i.e., the moment of an operator, $\expect{\hat{A}}$, can involve higher moments $\expect{\hat{B}\hat{C}}$ so that the equations do not close. In order to obtain closed equations that can be solved, we perform a cumulant approximation \cite{Kubo1962}. The first three cumulants are given by
\begin{align}
\cumulant{\hat{A}} &= \expect{\hat{A}}
\\
\cumulant{\hat{A}\hat{B}} &= \expect{\hat{A}\hat{B}} - \expect{\hat{A}}\expect{\hat{B}}
\\
\cumulant{\hat{A}\hat{B}\hat{C}} &= \expect{\hat{A}\hat{B}\hat{C}} - \expect{\hat{A}\hat{B}} \expect{\hat{C}} - \expect{\hat{A}\hat{C}} \expect{\hat{B}} - \expect{\hat{B}\hat{C}} \expect{\hat{A}} + 2 \expect{\hat{A}} \expect{\hat{B}} \expect{\hat{C}}.
\end{align}
Performing a cumulant approximation to lowest (first) order means that the second cumulant $\cumulant{\hat{A}\hat{B}} = 0$. This results in a factorization of the expectation value of any product of operators,
\begin{align}
\expect{\hat{A}\hat{B}} &= \expect{\hat{A}}\expect{\hat{B}}.
\end{align}
Such an approximation neglects correlations,
\begin{align}
\mathrm{corr}( \hat{A}, \hat{B} ) &= \frac{ \cumulant{ \hat{A} \hat{B} } }{ \expect{ \hat{A} } \expect{ \hat{B} } } = \frac{ \expect{ \hat{A} \hat{B} } - \expect{ \hat{A} } \expect{ \hat{B} } }{ \expect{ \hat{A} } \expect{ \hat{B} } }.
\label{eq:correlation}
\end{align}
We use this ansatz to derive a dynamical mean-field model for comparison in Sec. \ref{app:meanfield}. As this approximation misses out important terms, e.g., the dynamics of the oscillator population $\expect{\hat{n}} = \expect{\hat{a}^\dagger\hat{a}}$, as well as correlations, we perform a cumulant expansion to second order. Approximating $\cumulant{\hat{A}\hat{B}\hat{C}} = 0$ leads to
\begin{align}
\expect{\hat{A}\hat{B}\hat{C}} = \expect{\hat{A}\hat{B}} \expect{\hat{C}} + \expect{\hat{A}\hat{C}} \expect{\hat{B}} + \expect{\hat{B}\hat{C}} \expect{\hat{A}} -  2 \expect{\hat{A}} \expect{\hat{B}} \expect{\hat{C}}.
\end{align}
The resulting second-order cumulant equations include correlations, and thus describe phenomena where they are essential, such as the onset of the breakdown phase transition (BPT). The restriction to second-order moments can be justified by the pairwise nature of the interactions in the system, such as those between the spins and the harmonic oscillator. Starting from an initially uncorrelated state, for moderate couplings such interactions are assumed to build up at most two-body correlations. Indeed, such terms are sufficient to identify the relevant phase boundaries, in particular the BPT. However, as we discuss in Sec. \ref{app:intermediate}, the description falls short in predicting the correct scaling of the boundary with the system size.
\\

To describe the dynamics, we choose the observables
\begin{align}
&\hat{q} = \hat{a}^\dagger + \hat{a}
\\
&\hat{p} = i ( \hat{a}^\dagger - \hat{a} )
\\
&\hat{\sigma}^{(j)}_k, ~~~ \text{for $j \in \{1,\ldots,N\}$ and $k \in \{\mathrm{x},\mathrm{y},\mathrm{z}\}$}
\\
&\hat{n} = \hat{a}^\dagger \hat{a}
\\
&\hat{r} = \mathrm{Re} (\hat{a}^2)
\\
&\hat{s} = \mathrm{Im} (\hat{a}^2),
\end{align}
which are Hermitian so that the resulting equations are real. To simplify the equations, we assume that the system starts in a $Z_2$-respecting state \cite{Kirton2016} and discard all moments which do not respect this symmetry.
With this, we determine the equations of motion:

To first order we have a single equation that describes the polarization,
\begin{align}
\partial_t \expect{\hat{\sigma}_z^{(j)}} &= 2 g \expect{\hat{q} \hat{\sigma}_y^{(j)}} - \gamma (\expect{\hat{\sigma}_z^{(j)}} + 1).
\label{eq:evolution:Z:app}
\end{align}
To second order, we have three equations governing the dynamics of the harmonic oscillator, in particular the number of oscillator excitations $\hat{n}$,
\begin{align}
\partial_t \expect{\hat{n}} &= - \kappa \expect{\hat{n}} - N g \expect{\hat{p} \hat{\sigma}_x^{(j)}}
\\
\partial_t \expect{\hat{r}} &= - \kappa \expect{\hat{r}} + 2 \omega \expect{\hat{s}} + N g \expect{\hat{p} \hat{\sigma}_x^{(j)}}
\\
\partial_t \expect{\hat{s}}&= - \kappa \expect{\hat{s}} - 2 \omega \expect{\hat{r}} - N g \expect{\hat{p} \hat{\sigma}_x^{(j)}}.
\end{align}
Note that we have added a jump operator $\hat{L}_\kappa = \sqrt{\kappa} \hat{a}$ to the master equation which describes oscillator loss at a rate $\kappa$. The moments involving spins and oscillator evolve according to
\begin{align}
\partial_t \expect{\hat{q} \hat{\sigma}_x^{(j)}} &= - (\kappa + \gamma/2) \expect{\hat{q} \hat{\sigma}_x^{(j)}} +\omega \expect{\hat{p} \hat{\sigma}_x^{(j)}} - \omega_0 \expect{\hat{q} \hat{\sigma}_y^{(j)}}
\\
\partial_t \expect{\hat{p} \hat{\sigma}_x^{(j)}} &= - (\kappa + \gamma/2) \expect{\hat{p} \hat{\sigma}_x^{(j)}} - \omega \expect{\hat{q} \hat{\sigma}_x^{(j)}} - \omega_0 \expect{\hat{p} \hat{\sigma}_y^{(j)}} - 2 g ((N - 1) \expect{\hat{\sigma}_x^{(j)} \hat{\sigma}_x^{(k)}} + 1)
\\
\partial_t \expect{\hat{q} \hat{\sigma}_y^{(j)}} &= - (\kappa + \gamma/2) \expect{\hat{q} \hat{\sigma}_y^{(j)}} +\omega \expect{\hat{p} \hat{\sigma}_y^{(j)}} + \omega_0 \expect{\hat{q} \hat{\sigma}_x^{(j)}} - 4 g ( \expect{\hat{n}} + \expect{\hat{r}} + 1/2 ) \expect{\hat{\sigma}_z^{(j)}}
\\
\partial_t \expect{\hat{p} \hat{\sigma}_y^{(j)}} &= - (\kappa + \gamma/2) \expect{\hat{p} \hat{\sigma}_y^{(j)}} - \omega \expect{\hat{q} \hat{\sigma}_y^{(j)}} + \omega_0 \expect{\hat{p} \hat{\sigma}_x^{(j)}} - 4 g \expect{\hat{s}} \expect{\hat{\sigma}_z^{(j)}} - 2 g (N - 1) \expect{\hat{\sigma}_x^{(j)} \hat{\sigma}_y^{(k)}}.
\end{align}
Finally, the spin-spin moments are determined by
\begin{align}
\partial_t \expect{\hat{\sigma}_x^{(j)} \hat{\sigma}_x^{(k)}} &= - \gamma \expect{\hat{\sigma}_x^{(j)} \hat{\sigma}_x^{(k)}} - 2 \omega_0 \expect{\hat{\sigma}_x^{(j)} \hat{\sigma}_y^{(k)}} 
\\
\partial_t \expect{\hat{\sigma}_y^{(j)} \hat{\sigma}_y^{(k)}} &= - \gamma \expect{\hat{\sigma}_y^{(j)} \hat{\sigma}_y^{(k)}} + 2 \omega_0 \expect{\hat{\sigma}_x^{(j)} \hat{\sigma}_y^{(k)}} - 4 g \expect{\hat{q} \hat{\sigma}_y^{(j)}} \expect{\hat{\sigma}_z^{(j)}}
\\
\partial_t \expect{\hat{\sigma}_z^{(j)} \hat{\sigma}_z^{(k)}} &= - 2 \gamma (\expect{\hat{\sigma}_z^{(j)} \hat{\sigma}_z^{(k)}} + \expect{\hat{\sigma}_z^{(k)}}) + 4 g \expect{\hat{q} \hat{\sigma}_y^{(j)}} \expect{\hat{\sigma}_z^{(j)}}
\\
\partial_t \expect{\hat{\sigma}_x^{(j)} \hat{\sigma}_y^{(k)}} &= - \gamma \expect{\hat{\sigma}_x^{(j)} \hat{\sigma}_y^{(k)}} + \omega_0 ( \expect{\hat{\sigma}_x^{(j)} \hat{\sigma}_x^{(k)}} - \expect{\hat{\sigma}_y^{(j)} \hat{\sigma}_y^{(k)}} ) - 2 g \expect{\hat{q} \hat{\sigma}_x^{(j)}} \expect{\hat{\sigma}_z^{(j)}}.
\label{eq:evolution:Cxy:app}
\end{align}
Equations \eqref{eq:evolution:Z:app}--\eqref{eq:evolution:Cxy:app} can be used to numerically simulate the evolution and to analytically solve for the steady state. Based on the assumption that all spins are identical, in the main part we drop the superscript $j$ for moments involving a single spin such as $\expect{\hat{\sigma}_x^{(j)}}$ and $\expect{\hat{q} \hat{\sigma}_x^{(j)}}$. For two-spin correlators which describe with different spins $j$ and $k$ we define $\expect{\hat{\sigma}_\alpha \hat{\sigma}_\beta} \equiv \expect{\hat{\sigma}_\alpha^{(j)} \hat{\sigma}_\beta^{(k)}}$.

\subsection{Analytical solution for the steady state}

We solve Eqs. \eqref{eq:evolution:Z:app} -- \eqref{eq:evolution:Cxy:app} for the steady state by setting the time derivatives on the left-hand side to zero. In the following, we assume $\kappa=0$ and defer the discussion of the role of oscillator decay (or cooling) to Sec. \ref{app:cooling}. We then obtain the steady state solutions for the polarization
\begin{align}
\expect{ \hat{ \sigma }_z }_\infty &= \frac{ 1 }{ 8 ( N - 1 ) g^2 } \left( \pm D - \left( \omega \omega_{0,\gamma} + 4 ( N - 1 ) g^2 \right) \right)
,
\end{align}
the harmonic oscillator observables,
\begin{align}
\expect{ \hat{n} }_\infty &= 
\mp \left( \frac{ \omega^2 + \omega_0 \omega_{0,\gamma} }{ 8 \omega \omega_{0,\gamma} } + \frac{ \omega_0 }{ 16 ( N - 1 ) \omega g^2 } - \frac{ \omega_0 }{ 16 \omega g^2 } \right) D + \frac{ \omega_0 \omega_{0, \gamma} }{ 16 ( N - 1 ) g^2 }
\\ \nonumber
& ~~~ + ( N - 1 ) \left( \frac{ \omega^2 + \omega_0 \omega_{0,\gamma} }{ \omega \omega_0 - g^2 } \frac{ \omega_0 }{ 2 \omega_{0,\gamma} } - \frac{ 2 \omega^2 + \omega_0 \omega_{0,\gamma} }{ 4 \omega \omega_{0,\gamma} }\right)
+ \frac{ \omega^2 + \omega_0 \omega_{0,\gamma} }{ \omega \omega_0 - g^2 } \left( \frac{ \omega_{0,\gamma} }{ 8 \omega_{0,\gamma} } \right) + \frac{ 4 \omega_0 }{ \omega } - \frac{ \omega_0 \omega_{0,\gamma} }{ g^2 } - \frac{ 1 }{ 2 } 
\\
\expect{ \hat{r} }_\infty &= \frac{ N }{ 16 ( N - 1 ) \omega g^2 } \left( \pm D - \left( \omega \omega_{0,\gamma} + 4 ( N - 1 ) g^2 \right) \right)
\\
\expect{ \hat{s} }_\infty &= 0
,
\end{align}
the spin-oscillator correlations,
\begin{align}
\expect{ \hat{q} \hat{ \sigma }_x }_\infty &= - \frac{ \omega_0 }{ 16 ( N - 1 ) g^2 } \left( \pm D - \left( \omega \omega_{0,\gamma} - 4 ( N - 1 ) g^2 \right) \right)
\\
\expect{ \hat{p} \hat{ \sigma }_x }_\infty &= 0
\\
\expect{ \hat{q} \hat{ \sigma }_y }_\infty &= \frac{ \gamma }{ 16 ( N - 1 ) g^2 } \left( \pm D - \left( \omega \omega_{0,\gamma} - 4 ( N - 1 ) g^2 \right) \right)
\\
\expect{ \hat{p} \hat{ \sigma }_y }_\infty &= - \frac{ g }{ \omega_0 }
,
\end{align}
and the spin-spin correlations,
\begin{align}
\expect{ \hat{ \sigma }_x^{(j)} \hat{ \sigma }_x^{(k)} }_\infty &= \frac{ 1 }{ 16 ( N - 1 )^2 g^4 } \left( ( \pm D - ( \omega \omega_{0,\gamma} + 4 ( N - 1 ) g^2 ) ) \omega \omega_0 + 8 ( N - 1 ) ( \omega \omega_0 - g^2 ) g^2 \right)
\\
\expect{ \hat{ \sigma }_y^{(j)} \hat{ \sigma }_y^{(k)} }_\infty &= \frac{ ( \gamma / 2 )^2 }{ 16 ( N - 1 )^2 \omega_0 g^4 } \left( ( \pm D - ( \omega \omega_{0,\gamma} + 4 ( N - 1 ) g^2 ) ) \omega \omega_0 + 8 ( N - 1 ) ( \omega \omega_0 - g^2 ) g^2 \right)
\\
\expect{ \hat{ \sigma }_x^{(j)} \hat{ \sigma }_y^{(k)} }_\infty &= - \frac{ ( \gamma / 2 ) }{ 16 ( N - 1 )^2 \omega_0 g^4 } \left( ( \pm D - ( \omega \omega_{0,\gamma} + 4 ( N - 1 ) g^2 ) ) \omega \omega_0 + 8 ( N - 1 ) ( \omega \omega_0 - g^2 ) g^2 \right)
\\
\expect{ \hat{ \sigma }_z^{(j)} \hat{ \sigma }_z^{(k)} }_\infty &= - \frac{ 1 }{ 32 ( N - 1 )^2 \omega_0 g^4 } \left( \pm ( \omega \omega_0 + 4 ( N - 1 ) g^2 ) \omega_0 D - 16 ( N - 1 )^2 \omega_0 g^4 + 8 ( N - 1 ) \omega_{0,\gamma} + \omega^2 \omega_0 \omega_{0,\gamma}^2 ) \right).
\end{align}
Here, we have introduced the shorthand notation $\omega_{0,\gamma} = ( \omega_0^2 + (\gamma/2)^2 ) / \omega_0$.
The solutions appear in pairs separated by a discriminant,
\begin{align}
D = \sqrt{ 16 ( N - 1 ) g^4  \frac{ \omega_{0,\gamma} }{ \omega_0 } + ( \omega \omega_{0,\gamma} - 4 ( N - 1 ) g^2 )^2
}.
\label{eq:discriminant}
\end{align}
We consider these solutions in the thermodynamic limit of many particles in Sec.  \ref{app:many}, for a single particle in Sec. \ref{app:single}, and in the intermediate regime in Sec.  \ref{app:intermediate}.

\subsubsection{Thermodynamic limit} \label{app:many}

For many particles $N \gg 1$, we have two solutions
\begin{align}
\expect{ \hat{\sigma}_z^{(j)} } &= 
\frac{ 1 }{ 8 N g^2 } \left( \pm D - 4 N g^2 - \omega \omega_{0, \gamma} \right)
\label{eq:solution:Z:many}
\\
\expect{ \hat{n} } / N &= 
\frac{ \omega_0 }{ 16 \omega N g^2 } \left( \pm D + 4 N g^2-\omega \omega_{0, \gamma} \right)
\\
\expect{ \hat{r} }_\infty &= \frac{ 1 }{ 16 \omega g^2 } \left( \pm D - 4 N g^2 - \omega \omega_{0,\gamma} \right)
\\
\expect{ \hat{s} } / N &= 0
\\
\expect{ \hat{q} \hat{\sigma}_\mathrm{x}^{(j)} } &= 
-\frac{ \omega_0 }{ 8 (N g^2)^{3/2} } \left( \pm D +4 N g^2-\omega \omega_{0, \gamma} \right)
\\
\expect{ \hat{p} \hat{\sigma}_\mathrm{x}^{(j)} } &= 0
\\
\expect{ \hat{q} \hat{\sigma}_\mathrm{y}^{(j)} } &= 
\frac{ \gamma }{ 16 (N g^2)^{3/2} }  \left( \pm D + 4 N g^2 - \omega \omega_{0, \gamma} \right)
\\
\expect{ \hat{p} \hat{\sigma}_\mathrm{y}^{(j)} } &= 
-\frac{ g }{ \omega_0 }
\\
\expect{\hat{\sigma}_\mathrm{x}^{(j)} \hat{\sigma}_\mathrm{x}^{(k)}} &= 
\frac{ \omega \omega_0 }{ 16 N^2 g^4 } \left( \pm D + 4 N g^2 - \omega  \omega_{0, \gamma} \right)
\\
\expect{\hat{\sigma}_\mathrm{x}^{(j)} \hat{\sigma}_\mathrm{y}^{(k)}} &= 
-\frac{\gamma }{ 32 N^2 g^4 } \omega \left( \pm D + 4 N g^2 - \omega  \omega_{0, \gamma} \right)
\\
\expect{\hat{\sigma}_\mathrm{y}^{(j)} \hat{\sigma}_\mathrm{y}^{(k)}} &= 
\frac{\omega \gamma^2 }{ 64 \omega_0 N^2 g^4 } \left( \pm D + 4 N g^2 - \omega  \omega_\mathrm{0,\gamma} \right)
\\
\expect{\hat{\sigma}_\mathrm{z}^{(j)} \hat{\sigma}_\mathrm{z}^{(k)}} &= 
\frac{ 1 }{ 32 N^2 g^4 } \left( \mp D \left( 4 N g^2 + \omega \omega_{0, \gamma} \right) + 16 N^2 g^4 + \omega^2 \omega_{0, \gamma}^2 \right).
\label{eq:solution:Czz:many}
\end{align}
For the discriminant in Eq. \eqref{eq:discriminant} we have for $N \gg 1$,
\begin{align}
D = \sqrt{ \left( \omega \omega_{0,\gamma} - 4 N g^2 \right)^2 + 16 N g^4 \left( 1 + \frac{\gamma^2}{\omega_0^2} \right) }.
\label{eq:discriminant2}
\end{align}
Taking the derivative with respect to the coupling strength,
\begin{align}
\frac{\partial D}{\partial g} = \frac{1}{D^3} \sqrt{ 16 N g \left( 4 g^2 \left( N + \frac{(\gamma/2)^2}{\omega_0^2} \right) - \omega \omega_{0,\gamma} \right) },
\label{eq:discriminantderivative}
\end{align}
it can be seen that a non-analyticity can arise for $D=0$. From Eq. \eqref{eq:discriminant2}, it is apparent that such condition can only be reached in the thermodynamic limit, $N \rightarrow \infty$, where the term $(\gamma/2)^2/\omega_0^2$ is negligible compared to $N$. In this limit, the subleading term $\propto N g^4$ vanishes and the discriminant becomes
\begin{align}
D &\approx |\omega \omega_{0,\gamma} - 4 N g^2|.
\label{eq:discriminant3}
\end{align}
Thus, for $N \rightarrow \infty$ is it possible to have $D \rightarrow 0$. From Eq. \eqref{eq:discriminant3} we obtain the critical coupling $g_\mathrm{c} = \sqrt{ \omega \omega_\mathrm{0,\gamma} / ( 4 N ) }$, for which indeed $\partial D / \partial g \rightarrow \infty$. This non-analyticity corresponds to the superradiant phase transition.

\subsubsection{Single particle} \label{app:single}

Solving Eqs. \eqref{eq:evolution:Z:app}--\eqref{eq:evolution:Cxy:app} for $N=1$ yields a single solution
\begin{align}
\expect{ \hat{\sigma}_z^{(j)} } &= 
\frac{ g^2 }{ \omega \omega_0 } - 1
\\
\expect{ \hat{n} } &= 
\frac{ \omega^2 + \omega_0 \omega_{0, \gamma} }{ 4 \left( \omega \omega_0 - g^2 \right)} - \frac{ 1 }{ 2 } \left( 1 + \frac{ g^2 }{ \omega^2 } \right)
\\
\expect{ \hat{r} } &= \frac{g^2}{2 \omega ^2}
\\
\expect{ \hat{s} } &= 0
\\
\expect{ \hat{q} \hat{\sigma}_\mathrm{x}^{(j)} } &= -\frac{ g }{ \omega }
\\
\expect{ \hat{p} \hat{\sigma}_\mathrm{x}^{(j)} } &= 0
\\
\expect{ \hat{q} \hat{\sigma}_\mathrm{x}^{(j)} } &= \frac{ \gamma g } {2 \omega \omega_0 }
\\
\expect{ \hat{p} \hat{\sigma}_\mathrm{y}^{(j)} } &= - \frac{ g }{ \omega_0 }
\\
\expect{\hat{\sigma}_\mathrm{x}^{(j)} \hat{\sigma}_\mathrm{x}^{(k)}} &=
\frac{ 2 g^2 \left( \omega \omega_0 - g^2 \right) }{ \omega^2 \omega_0 \omega_{0, \gamma} }
\\
\expect{\hat{\sigma}_\mathrm{y}^{(j)} \hat{\sigma}_\mathrm{y}^{(k)}} &=
\frac{ \gamma^2 g^2 \left( \omega \omega_0 - g^2 \right)}{ 2 \omega^2 \omega_0^3 \omega_{0, \gamma} }
\\
\expect{\hat{\sigma}_\mathrm{z}^{(j)} \hat{\sigma}_\mathrm{z}^{(k)}} &=
\frac{ \left( \omega \omega_0 - g^2 \right)^2 }{ \omega ^2 \omega_0^2 }
\\
\expect{\hat{\sigma}_\mathrm{x}^{(j)} \hat{\sigma}_\mathrm{y}^{(k)}} &=
- \frac{ \gamma  g^2 \left( \omega \omega_0 - g^2 \right) }{ \omega^2 \omega_0^2 \omega_{0, \gamma} }.
\end{align}
Thus, for $N = 1$ no superradiant phase transition takes place.
On the other hand, the mean oscillator population number $\expect{\hat{n}}$ is found to have a pole for $g_\mathrm{b,1} = \sqrt{ \omega \omega_0 }$ which marks the breakdown of the oscillator blockade.

\subsubsection{Intermediate regime and discrepancies of the methods}  \label{app:intermediate}

In the intermediate regime with $1 < N \ll \infty$, we have that $D > 0$ so that the non-analyticity seizes to exist and the superradiant phase transition turns into a crossover.

In contrast to that, the oscillator breakdown phase transition is present for all finite system sizes. The cumulant expansion yields a steady-state phonon number dependence $\expect{\hat{n}} \propto (\omega \omega_0 - g^2)^{-1}$ which has a pole at $g_\mathrm{b,1} = \sqrt{ \omega \omega_0 }$. The scaling of this second phase boundary, in particular in conjunction with the simultaneous vanishing of coherences at the second phase transition, appears to suggest a single-particle effect. This disagrees with the results from our microscopic description and our numerical simulations. The discrepancy is a result of the factorization into pairwise correlations in our second-order cumulant expansion. While the assumption that at most pairwise correlations are built up by the two-body couplings in the Hamiltonian is correct for the anti-JC coupling below the critical coupling of the BPT, it cannot account for the concatenated JC couplings which give rise to multipartite correlations and thus lead to cooperative shifts. While a cumulant expansion to higher order may yield more accurate results at considerably increased analytical complexity, here we use a combination of a microscopic description and numerical simulations to resolve the discrepancy.

\subsection{Comparison to mean-field}\label{app:meanfield}

We compare the above results obtained from our second-order cumulant expansion with mean-field theory. To this end, we derive the dynamical equations factorizing second-order moments of operators, $\expect{\hat{A} \hat{B}} = \expect{\hat{A}} \expect{\hat{B}}$. The resulting semi-classical mean-field equations of motion are
\begin{align}
\partial_t \expect{\hat{q}} &= - (\kappa/2) \expect{\hat{q}} - \omega \expect{\hat{p}}
\\
\partial_t \expect{\hat{p}} &= - \omega \expect{\hat{q}} - (\kappa/2) \expect{\hat{p}} - 2 N g \expect{\hat{\sigma}_x^{(j)}}
\\
\partial_t \expect{\hat{\sigma}_x^{(j)}} &= - (\gamma/2) \expect{\hat{\sigma}_x^{(j)}} - \omega_0 \expect{\hat{\sigma}_y^{(j)}} 
\\
\partial_t \expect{\hat{\sigma}_y^{(j)}} &= + \omega_0 \expect{\hat{\sigma}_x^{(j)}} - (\gamma/2) \expect{\hat{\sigma}_y^{(j)}} - 2 g \expect{\hat{q}} \expect{\hat{\sigma}_z^{(j)}}
\\
\partial_t \expect{\hat{\sigma}_z^{(j)}} &= + 2 g \expect{\hat{q}} \expect{\hat{\sigma}_y^{(j)}} - \gamma ( \expect{\hat{\sigma}_z^{(j)}} + 1 )
\end{align}
Solving for the steady state, we obtain one solution with $\expect{\hat{q}} = \expect{\hat{p}} = \expect{\hat{\sigma}_x^{(j)}} = \expect{\hat{\sigma}_y^{(j)}} = 0$ and $\expect{\hat{\sigma}_z^{(j)}} = -1$, and a pair of solutions (here given for $\kappa=0$),
\begin{align}
\expect{\hat{q}} &= \mp \frac{D}{\sqrt{2} g \omega }
\\
\expect{\hat{p}} &= 0
\\
\expect{\hat{\sigma}_x^{(j)}} &= \pm \frac{D}{2 \sqrt{2} g^2 N}
\\
\expect{\hat{\sigma}_y^{(j)}} &= \mp \frac{\gamma D}{4 \sqrt{2} g^2 N \omega_0}
\\
\expect{\hat{\sigma}_z^{(j)}} &= -\frac{\omega \omega_{0,\gamma} }{4 g^2 N}.
\end{align}
with a discriminant $D = | 4 N g^2 - \omega \omega_{0,\gamma} |$. At $D=0$, there is a bifurcation which marks the superradiant phase transition (SPT). From $D=0$ we obtain the well-known result for the critical coupling,
\begin{align}
g_\mathrm{c} = \sqrt{ \frac{\omega \omega_{0,\gamma}}{4 N} }.
\end{align}
The SPT in the open-system Dicke model is thus described by a semi-classical, mean-field model, in an accordance with previous findings \cite{Dimer2007, Kirton2016, Larson2017}. On the other hand, the breakdown phase transition (BPT) is not obtained from the above treatment. We thus conclude that for the BPT correlations (as defined in Eq. \eqref{eq:correlation}) are essential. These are captured by our second-order cumulant expansion, as well as by our microscopic analytical theory, and our numerical simulations of the quantum master equation.
\\
\\


\begin{figure*}
\centering
\includegraphics[width=17.2cm]{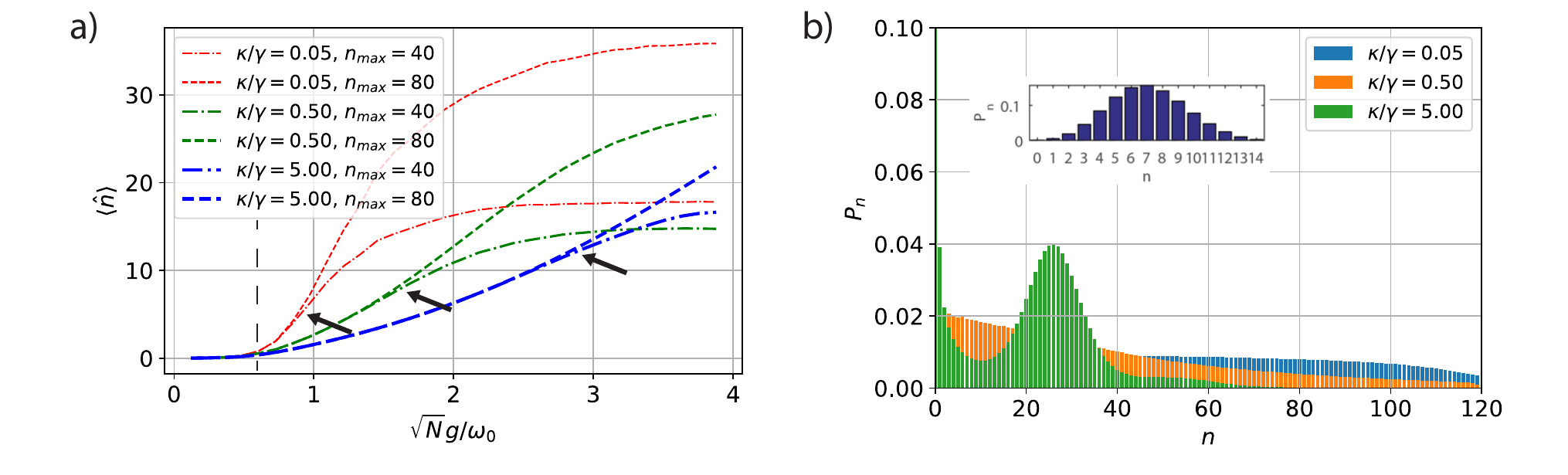}
\caption{
Effect of oscillator decay. (a) Oscillator decay or cooling of the oscillator at a rate $\kappa$ suppresses the breakdown phase transition (BPT) such that it occurs at higher coupling strength $g$. The BPT is visible from the onset of the cut-off dependence on the number of simulated oscillator quanta $n_\mathrm{max}$, and is marked by a black arrow. (b) Oscillator decay allows for the observation of the superradiant phase transition (critical coupling marked by a dashed line) with an oscillator distribution similar to the closed-system case (inset). Results are shown for $N=2$. Populations are sampled at $g=3.8 \omega_0$.
}
\label{fig:cooling}
\end{figure*}

\subsection{Effect of oscillator decay} \label{app:cooling}

For $\kappa > 0$, it is still possible to analytically solve the cumulant equations \eqref{eq:evolution:Z:app}--\eqref{eq:evolution:Cxy:app} for the steady state.
The resulting expressions become very lengthy and are therefore not displayed here.
Formally, there always exists one physical solution which implies that no breakdown occurs for $\kappa > 0$. However, our numerical simulations yield a different result, namely that the breakdown is shifted to higher coupling values as $\kappa$ is increased, as can be seen from Fig. \ref{fig:cooling}:
\\
For small cooling rates $\kappa$, the critical coupling for the BPT remains at values $g_\mathrm{b} \approx 2 g_\mathrm{c}$. As the cooling rate is increased, the breakdown is shifted to higher couplings $g_\mathrm{b} \gg g_\mathrm{c}$, way above the SPT. Hence, for sufficiently large cooling rates $\kappa$, the BPT is suppressed and the superradiant phase can be explored at in steady state also at higher couplings.

\section{Dressed-state excitation and criticality}
\label{app:dressed}

To illustrate the microscopic mechanism giving rise to the breakdown of the oscillator blockade, we consider the dressed states of the system. Since we assume the balanced Dicke model with $\omega_0 \approx \omega$, the Jaynes-Cummings (JC) Hamiltonian,
\begin{align}
\hat{H}_\mathrm{JC}
&= g \sum_{j=1}^N ( \hat{a}^\dagger \sigma_-^{(j)} + \hat{a} \sigma_+^{(j)} ),
\label{eq:JC_hamiltonian}
\end{align}
resonantly couples states with the same number of excitations, $n_\mathrm{ex}$, as is illustrated in Fig. 3 in the main part. It thereby hybridizes these levels into dressed states, which are probed by the anti-Jaynes-Cummings (anti-JC) Hamiltonian,
\begin{align}
\hat{H}_\mathrm{anti-JC}
&= g_+ \sum_{j=1}^N ( \hat{a}^\dagger \sigma_+^{(j)} + \hat{a} \sigma_-^{(j)} ).
\label{eq:anti_JC_hamiltonian}
\end{align}
This probe is a priori off-resonant, since creating two excitations, one to the spins and one to the oscillator, requires adding an energy of $\omega + \omega_0$. In the following, we diagonalize the Hamiltonian governing the JC-coupled subspaces,
\begin{align}
\hat{H}_{n_\mathrm{ex}} = \frac{\omega_0}{2} \sum_{j=1}^N \hat{\sigma}_z^{(j)} + \omega \hat{a}^\dagger \hat{a} + \hat{H}_\mathrm{JC}.
\label{eq:Hnex}
\end{align}
We regard the lower-excitation manifolds analytically, and then proceed using numerics to diagonalize higher ones. In our analytical treatment, we use the fact that the Hamiltonian is permutation-symmetric and work with the Dicke states,
\begin{align}
\ket{n} = \left( \sum_{m=1}^N \sigma_+^{(m)} \right)^n \ket{0}^{\otimes N}
\end{align}
In the zero-excitation manifold with $n_\mathrm{ex}=0$, there is only a single, uncoupled state,
\begin{align}
\ket{0, 0} = \ket{0}^{\otimes N} \ket{0},
\end{align} 
which resides at an energy $-N\omega_0/2$. For simplicity, we express the energies with respect to this value, thus assigning $\lambda_0=0$ to $\ket{0}\ket{0}$.
The second manifold, with $n_\mathrm{ex} = 1$, is described by the Hamiltonian
\begin{align}
H_1 = \omega \ket{0, 1} \bra{0, 1} + \omega_0 \ket{1, 0} \bra{1, 0} + \sqrt{N} g ( \ket{1, 0} \bra{0, 1} + \ket{0, 1} \bra{1, 0} ).
\end{align} 
The two coupled states, $\ket{1}\ket{0}$ and $\ket{0}\ket{1}$, are hybridized by $H_\mathrm{JC}$. Diagonalizing $H_1$, we obtain two dressed states,
\begin{align}
\ket{\pm} = \frac{1}{2} ( \ket{1} \ket{0} \pm \ket{0} \ket{1} ).
\end{align} 
For these states we obtain eigenenergies,
\begin{align}
\lambda_{1,\pm}
&= \frac{1}{2}\left(\omega + \omega_0 \pm \sqrt{(\omega + \omega_0)^2 - 4 ( \omega \omega_0 - N g^2 )}\right).
\end{align}
which are centered around and repelled from the mean energy of a single spin or oscillator excitation. Based on the assumption that $(\omega \omega_0 - 4 N g^2)^2 \ll (\omega + \omega_0)^2$, we can expand the square-root and obtain
\begin{align}
\lambda_{1,+}
&\approx \omega + \omega_0 - \frac{ \omega \omega_0 - N g^2 }{ \omega + \omega_0 },
\\
\lambda_{1,-}
&\approx \frac{ \omega \omega_0 - N g^2 }{ \omega + \omega_0 },
\end{align}
The next higher manifold with $n_\mathrm{ex}=2$ is excited from  $\ket{0,0}$ by the anti-JC coupling to $\ket{1,1}$ is described by
\begin{align}
H_2 = &2 \omega \ket{0, 2} \bra{0, 2} + (\omega + \omega_0) \ket{1, 1} \bra{1, 1} + 2 \omega_0 \ket{2, 0} \bra{2, 0}
\\
\nonumber
&+  \sqrt{2} g ( \sqrt{N} \ket{1, 1} \bra{0, 2} + \sqrt{N-1} \ket{1, 1} \bra{2, 0} + H.c.).
\end{align}
In the following, we assume a large number of spins $N \gg 1$ and thereby approximate $N - 1 \approx N$ and so on. We obtain for the eigenenergies in $n_\mathrm{ex}=2$-manifold,
\begin{align}
\lambda_{2,0} &= \omega + \omega_0
\\
\lambda_{2,\pm}
&= \left(\omega + \omega_0 \pm \sqrt{(\omega + \omega_0)^2 - 4 ( \omega  \omega_0 - N g^2 )}\right)
\end{align}
Here, $\lambda_{2,0}$ corresponds to a dark state $\ket{\psi_{2,0}}=(\ket{0, 2} - \ket{2, 0})/\sqrt{2}$ which is not excited from $\ket{0}\ket{0}$, while the states corresponding to $\lambda_{2,\pm}$ are coupled to $\ket{0}{0}$ by the anti-JC terms. Expanding the square-root, we obtain for the lower dressed-state energy of the two-excitation manifold,
\begin{align}
\lambda_{2,-}
&\approx \frac{ 2 ( \omega \omega_0 - N g^2 ) }{ \omega + \omega_0 }.
\end{align}
Higher manifolds are found to exhibit lowest dressed states with eigenvalues
\begin{align}
\lambda_{n_\mathrm{ex},-}
&= n_\mathrm{ex} \left( \frac{\omega + \omega_0}{2} - \sqrt{\left(\frac{\omega + \omega_0}{2}\right)^2 - ( \omega \omega_0 - N g^2 ) } \right)
\approx \frac{ n_\mathrm{ex} ( \omega \omega_0 - N g^2 ) }{ \omega + \omega_0 }.
%
\end{align}
The detuning between the maximally detuned dressed states of those manifolds which are coupled by the anti-JC terms, $n_\mathrm{ex}$ and $n_\mathrm{ex}+2$, is thus
\begin{align}
\Delta_{n_\mathrm{ex}} = \lambda_{n_\mathrm{ex}+2,\pm} - \lambda_{n_\mathrm{ex},\pm}
&\approx \frac{ 2 \left( \omega \omega_0 - N g^2 \right)} { \omega + \omega_0 },
\end{align}
where in the last step we have used that $\omega \approx \omega_0$. Similar resonance conditions are found for transitions between lesser detuned dressed states, as is demonstrated numerically below.
\\
Setting the coupling strength to $g_\mathrm{b} = \sqrt{\omega \omega_0 / N }$ thus brings the excitation by the anti-JC coupling in resonance. The combination with subsequent spontaneous emission leads to a net excitation of the harmonic oscillator. Subsequent excitation gives rise to the runaway of the oscillator population.

\begin{figure*}
\centering
\includegraphics[width=17.2cm]{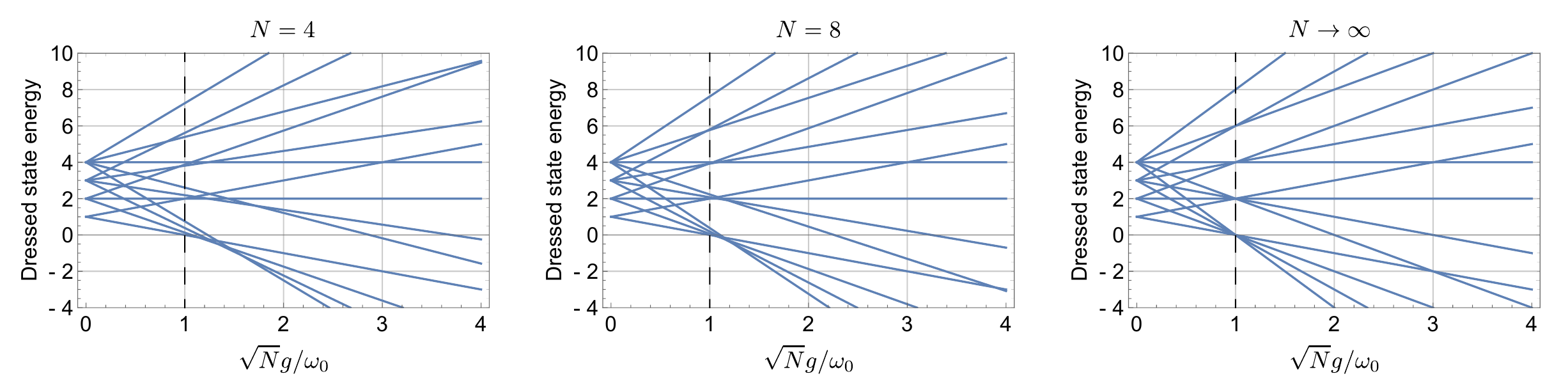}
\caption{
Dressed states and criticality. We plot the energy of the dressed states of the Hamiltonian in Eq. \eqref{eq:Hnex} for different numbers of spins $N=4,8$, as well as in the thermodynamic limit, $N \rightarrow \infty$, which are excited by the anti-JC Hamiltonian in Eq. \eqref{eq:anti_JC_hamiltonian}. It can be seen that for increasing particle number the resonances sharpen around the critical coupling of the breakdown phase transition, $g_\mathrm{b} = \sqrt{\omega \omega_0 / N} \approx 2 g_\mathrm{c}$ (marked by a dashed line). This confirms the cooperative and the critical nature of the BPT.
}
\label{fig:dressed}
\end{figure*}

\subsection{Criticality}

We demonstrate the criticality of the phenomenon numerically by showing the convergence of the resonance conditions corresponding to the excitation of the dressed states of of the system.

In Fig. \ref{fig:dressed}, we plot the energy of the dressed states of the subspaces coupled by the Hamiltonian in Eq. \eqref{eq:Hnex} for different numbers of spins. These eigenenergies set the resonance conditions for the excitation through the anti-JC coupling in Eq. \eqref{eq:anti_JC_hamiltonian}.

We calculate the eigenenergies for $N=4,8$, as well as in the thermodynamic limit, $N \rightarrow \infty$. It can clearly be seen that the resonance conditions converge around the previously calculated value for the critical coupling of the breakdown phase transition, $g_\mathrm{b} = \sqrt{\omega \omega_0 / N}$. The anti-JC excitation of dressed states, adding two excitations, thus becomes critical in the thermodynamic limit. Subsequent decay through spontaneous emission removes one excitation, leaving a net gain of one excitation in the harmonic oscillator. The coincidence of the resonance conditions thus leads to a breakdown of the oscillator blockade at different numbers of excitations, $n_\mathrm{ex}$, giving rise to the runaway of the oscillator population. The cooperative nature of the breakdown is evident from the agreement of the critical coupling (dashed line) for different numbers of spins. We thus confirm the cooperative and critical behavior of the breakdown of the oscillator blockade.
\\
\\


\begin{figure}
\centering
\includegraphics[width=8.6cm]{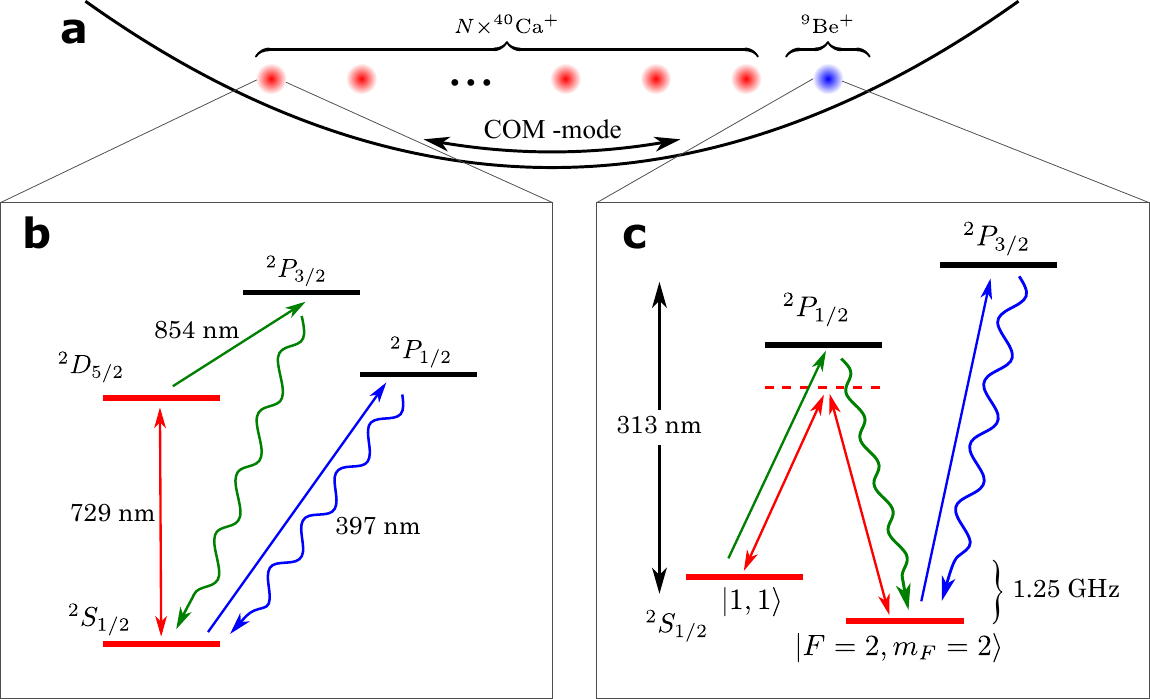}
\caption{Implementation of the Dicke model with trapped ions. (a) Dual species ions of calcium and beryllium are cotrapped and coupled to the COM mode of motion. The Dicke Hamiltonian with spontaneous emission is realized on $^{40}$Ca$^+$ while phonon dissipation is on $^9$Be$^+$. Relevant energy levels are shown for (b) $^{40}$Ca$^+$ and (c) $^{9}$Be$^+$.}
\label{fig:implement}
\end{figure}

\section{Implementation} \label{app:implementation}

We consider an implementation of the model using a system of trapped ions. 
To this end, we generalize the recipes in \cite{Pedernales2015, Lv2018, Aedo2018}. We consider a linear Paul trap coupled to the center-of-mass (COM) mode of motion, where all the ions are oscillating in phase (Fig. \ref{fig:implement} a)). In the low-energy regime, the system is a perfect approximation of $N$ spins interacting with a harmonic oscillator, where the spin is encoded in the ion's internal states.

The non-interacting Hamiltonian is given by
\begin{equation}
\hat{H}_0 = \frac{\omega_a}{2} \sum_j^N \hat{\sigma}_z^{(j)} + \omega_m \hat{a}^\dagger \hat{a},
\end{equation}
where $\omega_a$ and $\omega_m$ are the atomic transition and motional frequencies, respectively. A bichromatic laser field of frequencies $\omega_{r,b}$ drives at the same time both red and blue sideband interactions. We assume a small detuning from the red (blue)-sideband $\delta_r = \omega_a - \omega_m - \omega_r$ ($\delta_b = \omega_a + \omega_m - \omega_b)$.
The interaction Hamiltonian in the Lamb-Dicke regime reads
\begin{equation}
\hat{H}_\mathrm{int} = \eta \Omega_r\sum_{j=1}^N\hat{\sigma}_+^{(j)}\hat{a}e^{i\delta_r t} + \eta \Omega_b\sum_{j=1}^N\hat{\sigma}_+^{(j)}\hat{a}^\dagger e^{i\delta_b t} + H.c.
\end{equation}
Here $\Omega_{r,b}$ are the Rabi frequencies and $\eta$ is the Lamb-Dicke parameter. Changing to a rotating frame where $H_\mathrm{int}$ becomes time independent yields the effective Hamiltonian
\begin{align}
H_\mathrm{eff} 
&= \frac{\delta_\mathrm{r} + \delta_\mathrm{b}}{4} \sum_{j=1}^N \hat{\sigma}_z^{(j)} + \frac{\delta_\mathrm{b} - \delta_\mathrm{r}}{2} \hat{a}^\dagger \hat{a} + g ( \hat{a}^\dagger + \hat{a} ) ( \hat{\sigma}_+^{(j)} + \hat{\sigma}_-^{(j)} )
\\
&\equiv \omega \hat{a}^\dagger \hat{a} + \frac{\omega_0}{2} \sum_{j=1}^N \hat{\sigma}_z^{(j)} + g \sum_{j=1}^N ( \hat{a}^\dagger + \hat{a} ) \hat{\sigma}_x^{(j)}
\end{align}
which takes the form of the Dicke Hamiltonian (1), where we relate $\omega_0 = \frac{\delta_b+\delta_r}{4}$, $\omega = \frac{\delta_r-\delta_b}{2}$ and $g = g = \eta \Omega$.

Different interaction directions, e.g., along the $z$-direction \cite{Genway2014, Safavi2018}, as well as differences in the trap geometry (Paul vs. Penning trap) or transition (optical vs. hyperfine transition) allow for the implementation of similar spin-boson models.
\\

In addition to these couplings of the Dicke Hamiltonian, we engineer controlled spontaneous emission. For this purpose we consider exclusively calcium ions $^{40}$Ca$^+$. However the concept and technique can be easily extended to other ion species. Figure \ref{fig:implement} b) sketches the energy levels of  $^{40}$Ca$^+$, where we choose $\ket{^2S_{1/2}, m_J = 1/2}$ and $\ket{^2D_{5/2}, m_J = 3/2}$ as spin $\ket{0}$ and $\ket{1}$ states, respectively. The two states are connected by a quadrupole transition at 729 nm. Since the natural lifetime of the $\ket{^2D_{5/2}}$ state is longer than 1 s, effective spin decay is obtained by optical pumping, coupling $\ket{1}$ to the short-lived $\ket{^2P_{3/2}} \equiv \ket{e}$ state (lifetime $\sim 7$ ns) by a 854 nm ``repumper'' laser, modeled by a Hamiltonian
\begin{align}
\hat{H}_\mathrm{rep} = \frac{\Omega_\mathrm{rep}}{2} \sum_{j=1}^N \left( \ket{e}_j \bra{1} + \ket{1}_j \bra{e} \right).
\end{align}
$\ket{^2P_{3/2}}$ decays back to $\ket{0}$ at a rate $\Gamma_{e0}$, as described by the jump operator
\begin{align}
\hat{L}_\mathrm{\Gamma_{e0}}^{(j)} = \sqrt{\Gamma_{e0}} \ket{0}_j \bra{e}.
\end{align}
For a total decay rate $\Gamma_e$ from $\ket{e}$ with $\Gamma_e^2 \gg \Omega_\mathrm{rep}^2$, we can use the effective operator formalism to adiabatically eliminate $\ket{e}$. We thereby obtain the effective jump operator
\begin{align}
\hat{L}_\mathrm{\Gamma_{0e},eff}^{(j)} = \sqrt{\frac{\Gamma_{0e} \Omega_\mathrm{rep}^2}{\Gamma_e^2}} \ket{0}_j \bra{1} \equiv \sqrt{\Gamma_{0e,  \mathrm{eff}}} \sigma_-^{(j)},
\end{align}
describing an incoherent decay from $\ket{1}$ to $\ket{0}$ with an effective decay rate $\Gamma_{0e, \mathrm{eff}}$. Associating $\gamma \equiv \Gamma_{0e, \mathrm{eff}}$ we obtain the jump operators of Eq. (2).
This effective decay rate is tunable by varying the Rabi frequency $\Omega_\mathrm{rep}$ of the 854 nm light.

All relevant parameters of the Dicke Hamiltonian and spontaneous emission are thus determined by the power and detuning of lasers, which are experimentally controlled and can be freely tuned. Additional decay channels opening up to levels other than the spin states need to be counteracted by  repumpers. These can be modeled in the same manner. Operating them at large  rate helps to minimize the population outside the desired Hilbert space.

\subsection{Observability of the oscillator population}

To characterize the breakdown phase transition it is essential to have access to the phonon Fock state distribution. For that, a traditional technique for phonon diagnosis can be applied, recording a blue-sideband flopping which maps the phonon distribution onto the spectrum of the ion oscillations. The internal state of $^{40}$Ca$^+$ is read out by fluorescent imaging on the $\ket{^2S_{1/2}}$ and $\ket{^2P_{1/2}}$ transition at 397 nm.

\subsection{Cooling}

To investigate the role of oscillator decay or to suppress the breakdown, it is possible to open a phonon loss channel without adding any significant disturbance on the Dicke Hamiltonian. This can be achieved either by sympathetic cooling on another co-trapped ion species, e.g., beryllium ions as depicted in Fig. \ref{fig:implement} or by performing local cooling on extra $^{40}$Ca$^+$ ion(s) with the help of single-ion addressing. Either method will multiply the complexity of the experimental design and setup, but has been successfully implemented by many ion trapping experiments in other contexts \cite{Ballance2015, Negnevitsky2018}. Take as an example the case of $^9$Be$^+$, all relevant transitions are at around 313 nm. Sideband cooling to near the motional ground state is performed by a combination of Raman stimulated driving and optical pumping on the $^2S_{1/2}$, $|F=2, m_F=2\rangle$ and $|F=1, m_F=1\rangle$ transition, mediated by the $^2P_{1/2}$ state. Here a cooling rate of a few tens of kHz can be expected. For a stronger cooling rate in the hundred kHz range, one can apply Doppler cooling on the closed cycling transition between $^2S_{1/2}$ $|F=2, m_F=2\rangle$ and $^2P_{3/2}$ $|F=3, m_F=3\rangle$ \cite{Meekhof1996, Kienzler2015}.

\subsection{Imperfections}

Decay from $\ket{e}$ to $\ket{1}$ results in an effective dephasing process which does not affect our observations. On the other hand, leakage out of the Hilbert space comprising $\{\ket{0},\ket{1}\}$ has to be avoided or countered by additional repump fields. Residual coupling of the sideband lasers to the carrier transition $\ket{0} \leftrightarrow \ket{1}$ is not found to have an effect.

\end{document}